\documentclass[prb,showpacs,preprintnumbers,amsfonts,amssymb,amsmath,floats,twocolumn,aps]{revtex4}

\usepackage{graphicx}
\usepackage{dcolumn}
\usepackage{bm}
\usepackage{color}
\usepackage[hyperindex,pdfmark]{hyperref}

\definecolor{red}{rgb}{1,0,0}
\newcommand{\beeq}{\begin{equation}}
\newcommand{\eneq}{\end{equation}}
\newcommand{\beeqa}{\begin{eqnarray}}
\newcommand{\eneqa}{\end{eqnarray}}
\newcommand{\vc}[1]{\mathbf #1}
\newcommand{\Tr}{\mbox{Tr}}
\newcommand{\tr}{\mbox{tr}}
\newcommand{\Real}{\mbox{Re}}
\newcommand{\Img}{\mbox{Im}}

\begin{document}
	\preprint{APS/123-QED}
	\title{Quantitative aspects of the dynamical impurity approach} 
	\author{Krunoslav Po\v{z}gaj\v{c}i\'{c}} \email{kpozga@lusi.uni-sb.de}
	\affiliation{Universit\"{a}t des Saarlandes, 
				Institut f\"{u}r Theoretische Physik\\
				Saarbr\"ucken, Germany}
	\date{\today}
%
%
%
\begin{abstract}												
 A calculation technique in the context of the self-energy functional 
 approach (SFA) and its local form, the dynamical impurity approach
 (DIA)\cite{Potthoff0103}, will be proposed. This technique allows for a
 precise calculation of the derivatives of the grand potential functional
 used in the search for a stationary point. To make a closer comparison
 of the DIA with the dynamical mean-field theory (DMFT)\cite{GKKR}, and to
 demonstrate the proposed technique, we calculated paramagnetic U-T phase
 diagram of the Hubbard model at half-filling, which exhibits 
 metal-insulator transition.
\end{abstract}															
%
%


	\maketitle
%
%
%
%
%
%
%
%
%
\section{\label{sec:intro}Introduction}
 Variational determination of the grand potential functional has a long
history. The seminal paper, made in the 1960s  
\cite{Lutt_Ward}, provides us with the expression for the
grand potential functional which reads
%
%
\beeqa                  %
\Omega[\vc{G}]=\Phi[\vc{G}]+\Tr\ln(-\vc{G})-\Tr((\vc{G}_0^{-1}-\vc{G}^{-1})\vc{G}),
\label{eq:Baym-Kadanoff}
\eneqa                  %
%
 where $\vc{G}_0$ is the free Green's function, $\vc{G}$ is the full Green's
function and $\Phi[\vc{G}]$ is the Luttinger-Ward functional expressed by a sum
over all connected skeleton diagrams. $\Omega[\vc{G}]$ is stationary for the physical
Green's function $\vc{G}$. The main problem which arises in that approach is
a formidable skeleton perturbation expansion.
 Approaches which used Eq.~(\ref{eq:Baym-Kadanoff}) as a starting point 
would approximate the Luttinger-Ward functional by an incomplete
diagram series or by composing $\Phi[\vc{G}]$ from only a few lowest order
diagrams\cite{Baym0862}.
 Recently it has been proposed to regard the grand potential
functional as a functional of the self-energy with the Legendre transform of
the Luttinger-Ward functional calculated from a simplified Hamiltonian
(which we will refer to as a reference system) and the self-energy of the
original system approximated by the one of the reference
system\cite{Potthoff0103}.
Mathematically expressed, the proposal states that
%
%
%
\beeqa                   %
\Omega_{\bf t}[{\bf\Sigma}({\bf t}')]&=&\Omega_{{\bf t}'}[{\bf\Sigma}({\bf
t}')]
+\Tr\ln(-({\bf G}_0^{-1}-{\bf \Sigma}({\bf t'}))^{-1})
\nonumber\\
&-&\Tr\ln(-({\bf G'}_0^{-1}-{\bf \Sigma}({\bf t'}))^{-1}),
\label{eq:thermo_potential}
\eneqa                   %
%
%
%
 where $\vc{\Sigma}[\vc{t'}]$ is the exact self-energy of the reference
system, $\vc{G_0'}$ is the free Green's function of the same and $\vc{G_0}$
is the free Green's function of the original system. The reference Hamiltonian
is chosen such that it is simpler than the original one, while a
systematic increase in its size would reproduce the original system. In
the case of the Hubbard model
\cite{Anderson1959,Kanamori1963,Hubbard1963}
, correlated sites
in the reference system correspond to those from the original one.
We can add uncorrelated sites to correlated sites
to mimic the
rest of the lattice\cite{Potthoff0103,Potthoff0603}. The grand potential
functional is stationary with respect to the variation of the exact
self-energy. The same stationarity condition will be imposed to the
approximate one defined by Eq.~(\ref{eq:thermo_potential}) with the
reference system being simpler than the original one.
 The simplest form of the reference system is obtained if we restrict our
attention to the local self-energy, in which case we speak of the dynamical
impurity approach (DIA). 
 In general, including reference systems with
non-local self-energies, the method is called self-energy functional
approach.
 In this account we will concentrate on the calculation performed in the
context of DIA, even though, only slight changes are necessary for a more
general, SFA approach.
\par
 In the seminal paper\cite{Potthoff0103} it was demonstrated on the
example of the Mott-Hubbard transition how even a rudimentary reference
system can give a good account of the physics obtained by the numerically
more expensive DMFT approach. We are going to address the question of how
the extension of the reference system improves the results. Since DIA is
equivalent to DMFT only when the number of uncorrelated sites goes to
infinity, it will be interesting to compare the convergence of the results
obtained in DIA to those obtained in DMFT as the number of uncorrelated
sites in the reference system is increased.
 Contrary to the mentioned rudimentary case with only one
variational variable, the set of variational parameters in the reference system
with six atoms has five components for the half-filled case. To be able to determine
borders of the phase diagram precisely, we calculate derivatives of
the grand potential functional analytically. This procedure will be sketched
in Section \ref{sec:stat_point} together with additional calculations
given in the appendices. In the Section \ref{sec:MI_Hubbard} we present the
results for the Hubbard model.
%
%
%
%
%
%
\section{Determination of the stationary point\label{sec:stat_point}}
 The main numerical burden of the SFA/DIA method is the determination of the
self-energy. That part consists of the diagonalization of the reference
system, calculation of the Green's function from the Lehmann representation
and, using Dyson's equation, determination of the self-energy. The only
numerically problematic part is the one of the calculation of the
self-energy from the Green's function (see Appendix \ref{sec:Self_energy}).
With the obtained self-energy we can evaluate trace terms along the lines
already discussed in some earlier accounts\cite{Potthoff0603}. That approach
would demand numerical calculation of derivatives for the determination of the
stationary points of the grand potential functional. The numerical load for
a reliable and precise calculation of derivatives can be large\cite{Press}.
 We will show how the additional information from the
diagonalized reference system can be used for their determination. The
biggest advantage of the method can be seen in the coexistence region of the
phase diagram.
\par
 Before embarking on the calculation of derivatives of the trace terms, we
are going to sketch a few steps in the calculation of the trace terms
already discussed \cite{Potthoff0603}.
 Using a general form of the Green's function $\vc{G}$, the trace term can
be written as
%
%
\beeq					%
 \Tr\ln(-\vc{G})=\sum_{\omega_n}\frac{e^{i\omega_n0^+}}{\beta}{\rm tr}
\ln\frac{-1}{i\omega_n+\mu-{\bf t}-{\bf \Sigma}(i\omega_n)},
\eneq					%
%
%
 where ${\rm tr}$ denotes trace over all quantum states. A function of a
Hermitian operator is defined as $f(A)=Uf(a)U^\dagger$ where $a$ is the 
diagonalized $A$ matrix and $U$ is the unitary transformation which performs it.
If $\eta_k(i\omega_n)$ are eigenvalues of the operator ${\bf t}+{\bf \Sigma}(i\omega_n)$, 
then
%
%
\beeq					%
 \Tr\ln(-\vc{G})=\sum_{\omega_n}\sum_k\frac{e^{i\omega_n0^+}}{\beta}
\ln\frac{-1}{i\omega_n+\mu-\eta_k(i\omega_n)}.
\label{eq:w_rearrange}
\eneq					%
%
%
 Summation over $k$ encompasses all eigensolutions of the operator 
${\bf t}+{\bf \Sigma}(i\omega)$. In order to get over from the 
summation over Matsubara frequencies to the integration in
the complex plane, a standard trick is applied. It consists 
in the introduction of a function in the complex plane which has 
poles at $z=i\omega_n$ and the subsequent integration around the
contour which includes the imaginary axis
%
%
\beeq					%
 \Tr\ln(-\vc{G})=
\sum_k \oint_C \frac{-dz}{2\pi {\rm i}} \frac{e^{z 0^+}}{e^{\beta
z}+1}\ln\frac{-1}{z+\mu-\eta_k(z)}.
\label{eq:integral_Cauchy}
\eneq					%
%
%
%
 The integration contour in Eq.~(\ref{eq:integral_Cauchy}) can be mapped
to the integration contour enclosing the real axis by the lines
infinitesimally
above and below it ($\omega\pm i\delta$).
 Using analytic properties of the integrand, i.e., 
$\eta_k(\omega-i\delta)=\eta_k^{*}(\omega+i\delta)$, 
and retaining only the largest contributions in $0^+$ and $\delta$ (both are
infinitesimally small positive numbers), we can express $\Tr\ln(-\vc{G})$ as
%
%
\beeqa					%
\sum_k \int_{-\infty}^{\infty}\frac{-d\omega}{\pi}
f(\omega)\Img \ln\frac{-1}
			{\omega+i0^{+}+\mu-\eta_k(\omega+i 0^{+})},
\eneqa					%
%
%
 where $f(\omega)=1/(e^{\beta\omega}+1)$ is the Fermi-Dirac function.
 For a function  $F(\omega+i0^{+})$ that has the property 
$\Img F(\omega+i0^{+})\leq 0$ (that is the case for the diagonal Green's
function), there follows the equality $\Img\ln(-F(\omega+i0^{+}))=\pi\Theta(F(\omega))
=\pi\Theta(1/F(\omega))$, with $\Theta$ being the step function. Due to the
finite number of poles in the reference system, singular points don't
give a finite contribution to the integral. The resulting expression for the
trace term reads
%
%
%
\beeqa					%
 \Tr\ln(-\vc{G}) 
= -\sum_k\int_{-\infty}^{\infty}d\omega
f(\omega)\Theta\left(\omega+\mu-\eta_k(\omega)\right).
\label{eq:tr_step_func}
\eneqa					%
%
%
 This expression is our starting point for calculating derivatives with
respect to the parameters of the reference system. As in the calculation of
the grand potential functional\cite{Potthoff0603}, we want to obtain the
integrand as a sum of the Dirac $\delta$-functions, which is then trivially
integrated. For this purpose we just have to exchange integration over the
real axis and derivative with respect to the reference system parameter $\lambda$.
 Due to the discontinuity of the argument of the step function in
Eq.~(\ref{eq:tr_step_func}), we obtain two sums of the $\delta$-functions, one
evaluated in the zeros of $\omega+\mu-\eta_k(\vc{\Sigma}(\vc{t'}),\omega)$ and
the other evaluated in the poles of $\eta_k(\vc{\Sigma}(\vc{t'}),\omega)$.
\begin{widetext}
\beeqa					%
&&\hspace*{-1.0cm} \frac{\partial}{\partial \lambda} \Tr\ln(-\vc{G}[\vc{\Sigma}(\vc{t'})]) 
= -\frac{\partial}{\partial \lambda} \sum_k 
			\int_{-\infty}^{\infty}d\omega
f(\omega)\Theta\left(\omega+\mu-\eta_k(\vc{\Sigma}(\vc{t'}),\omega)\right)\\
&=&\sum_k	\int_{-\infty}^{\infty}d\omega f(\omega)
			\left[
			\delta(\omega+\mu-\eta_k(\omega))\frac{\partial \eta_k(\vc{\Sigma}(\vc{t'}),\omega)}{\partial\lambda}
			-
			  \delta\left(\frac{1}{\omega+\mu-\eta_k(\vc{\Sigma}(\vc{t'}),\omega)}\right)
			 \frac{\partial \frac{1}{\omega+\mu-\eta_k(\vc{\Sigma}(\vc{t'}),\omega)}}{\partial\lambda}
			\right]
			\nonumber\\
&=&\sum_{k,i} 
			\int_{-\infty}^{\infty}d\omega f(\omega)
			\left[
			\frac{\partial \eta_k(\vc{\Sigma}(\vc{t'}),\omega)}{\partial\lambda}
			\frac{\delta(\omega-\omega_{i,k}^{(z)})}{\left|1-\frac{\partial \eta_k(\vc{\Sigma}(\vc{t'}),\omega)}{\partial \omega}\right|}
			-
			\frac{\partial \left(\frac{1}{\omega+\mu-\eta_k(\vc{\Sigma}(\vc{t'}),\omega)}\right)}{\partial\lambda}
			\frac{\delta(\omega-\omega_{j,k}^{(p)})}
			{\left|\frac{\partial(\frac{1}{\omega+\mu-\eta_k(\vc{\Sigma}(\vc{t'}),\omega)})}{\partial \omega}
			\right|} 
			\right]\\
&=&\sum_{k,i} 
			f(\omega_{i,k}^{(z)})
			\frac{\partial \eta_k({\bf \Sigma}({\bf t'}),\omega_{i,k}^{(z)})}{\partial\lambda}
			\frac{1}{\left|1-\frac{\partial \eta_k(\vc{\Sigma}(\vc{t'}),\omega)}{\partial \omega}\right|_{\omega=\omega_{i,k}^{(z)}}}
-\sum_{k,j} 
			f(\omega_{j,k}^{(p)})
			\frac{\partial \eta_k({\bf \Sigma}({\bf t'}),\omega_{j,k}^{(p)})}{\partial\lambda}
			\frac{1}{\left|1-\frac{\partial \eta_k({\bf \Sigma}({\bf t'}),\omega)}{\partial
			\omega}\right|_{\omega=\omega_{j,k}^{(p)}}}.
\label{eq:tr_deriv}
\eneqa					%
\end{widetext}
 The second term in Eq.~(\ref{eq:tr_deriv}) should be thought of as an evaluation of the
derivatives for $\omega\to\omega_{j,k}^{(p)}$, since each of the derivatives
diverges for $\omega=\omega_{j,k}^{(p)}$. 
 Using the notation of Eqs.~(\ref{eq:sig_analytic}, \ref{eq:dsig_analytic}), but with coefficients being
matrices instead of scalars, we can obtain the final expression for the
derivative of the trace term. It reads
\begin{widetext}
\beeqa					%
&& \hspace*{-0.8cm}\frac{\partial}{\partial t_i} \Tr\ln(-G[\Sigma(\vc{t'})]) 
=-\sum_{k,j} 
            f(\omega_{j,k}^{(p)})
            \frac{\langle\Psi_{k,j}|\vc{C}_j^{(2),d\Sigma}|\Psi_{k,j}\rangle}
            {\left|\langle\Psi_{k,j}|\vc{C}_j^{(1),\Sigma}|\Psi_{k,j}\rangle\right|}
+\sum_{k,i} 
			f(\omega_{i,k}^{(z)})
			\frac{\partial \eta_k(\Sigma({\bf t'}),\omega_{i,k}^{(z)})}{\partial t_i}
			\frac{1}{\left|1-\frac{\partial \eta_k(\Sigma(\vc{t'}),\omega)}
			{\partial \omega}\right|_{\omega=\omega_{i,k}^{(z)}}},
\label{eq:Tr_deriv_end}
\eneqa					%
\end{widetext}
 where $|\Psi_{k,j}\rangle$ is an eigenstate of the matrix
$\vc{C}^{(1),\Sigma}_j$.
 Equation (\ref{eq:Tr_deriv_end}) is completely general, i.e., it can also be applied
to the case of non-local self-energy. From now on, we are going to apply it
only to the case of the local self-energy.
\par
 For obtaining the grand potential functional we have to evaluate two trace
terms: one for the Green's function of the original system and one for the
Green's function of the reference system [see Eq.~(\ref{eq:Baym-Kadanoff})]. 
 Reference systems with the local self-energy can have any form (chainlike,
starlike etc.), because reference systems with different configuration of
bath sites can be mapped to each other so that the physical Green's function
 doesn't change. We have chosen the star-like reference system for its
simplicity, with the Hamiltonian given as (the paramagnetic state is assumed)
%
%
\beeqa                  %
\label{eq:H_SIAM}
H^{SIAM}&&=\sum_{\sigma}(\epsilon_{c}-\mu)c_{\sigma}^\dagger
	c_{\sigma}+U n_{\uparrow}n_{\downarrow}\nonumber\\
	&&\hspace{-2.0cm}+\sum_{l,\sigma}(\epsilon_{a,l}-\mu) a_{l,\sigma}^\dagger a_{l,\sigma}
	+\sum_{l,\sigma}V_{l}(c^\dagger_{l,\sigma}a_{l,\sigma}
	+a^\dagger_{l,\sigma}c_{l,\sigma}).
\eneqa                  %
%
%
 The trace of the full Green's function of the reference system can
be decomposed in a number of trace terms with the scalar Green's
functions
%
\beeqa					%
{\tr}'\ln(-\vc{G}'(i\omega))&&\nonumber\\
&&\hspace{-2.5cm}=\sum_\sigma\ln(-G_{1,\sigma}'(i\omega))+\sum_\sigma\sum_{k=2}^{N_a}\ln(-G_{k,\sigma}'(i\omega)),
\eneqa					%
%
with
$G'_k(i\omega_n)=1/(i\omega_n+\mu-\epsilon_k)$
and
%
\beeqa					%
G'_1(i\omega_n)=\frac{1}{i\omega_n+\mu-\epsilon_1-\sum_{k=2}^{N_a}
				\frac{V_k^2}{i\omega_n+\mu-\epsilon_k}-\Sigma(i\omega_n)}.
\eneqa					%
%
 Using the notation of Eq.~(\ref{eq:tr_deriv}), the following
identifications can be done:
%
%
\beeqa					%
\eta_1(i\omega_n)=\epsilon_1+\sum_{k=2}^{N_a}
                \frac{V_k^2}{i\omega_n+\mu-\epsilon_k}+\Sigma(i\omega_n) \\
\eta_{k>1}=\epsilon_k
\eneqa					%
%
%
 The original Green's function is diagonalized by the Fourier transformation
under assumption of the local self-energy and translation invariance of
the system. In this case $\eta_{k,\sigma}(\omega)=\epsilon(k)+\Sigma(\omega)$,
where $\epsilon(k)$ is the Fourier transformed one-particle term and $k$ denotes 
a wave number.
\par
 Stationary point in the parameter space $\vc{t}^\prime$ assures us that
we have found the self-energy which is physical. Since it is possible to
find more than one solution, thermodynamically stable are only those
characterized by the minimum of the grand potential\cite{Potthoff0603}.
%
%
%
%
%
%
%
%
%
\section{\label{sec:MI_Hubbard}Metal-insulator transition in 
the Hubbard model}%
 As a demonstration of the above numerical procedure, we determined the U-T phase
diagram of the Hubbard model in the paramagnetic state.
 In addition to the results for two-site DIA which have shown
qualitative agreement with the full DMFT calculations, we will
demonstrate that richer reference systems allow us to make quantitative
comparisons with the existing DMFT calculations\cite{Tong2001,Joo,Bulla}.
\par
\begin{figure}[htb]
\includegraphics[height=6.0cm]{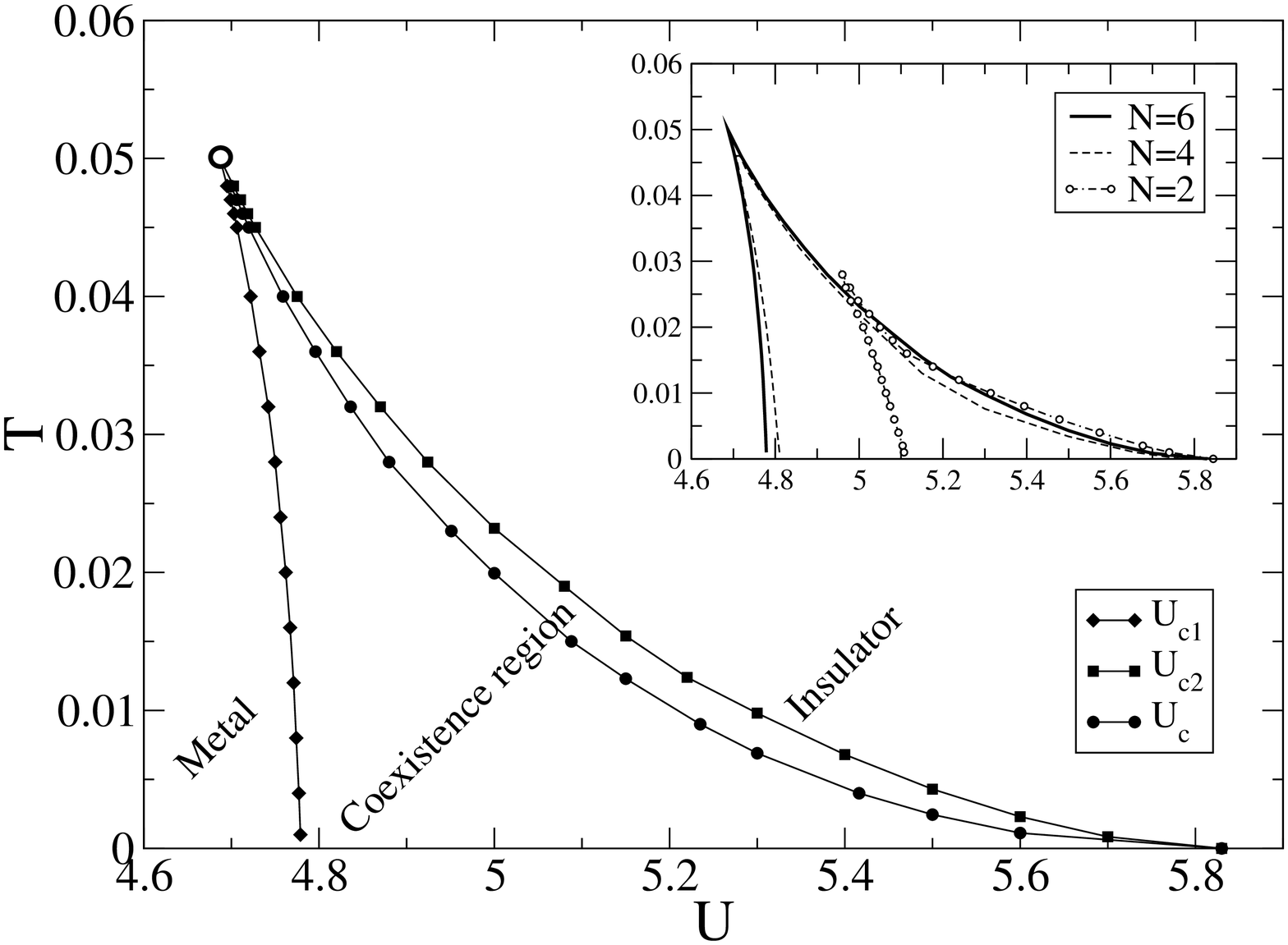}
\vspace*{0.3cm}
\caption{\label{fig:DIA_convergence}
 U-T phase diagram of the paramagnetic state of the Hubbard model. $U_{c1}$
curve denotes the left border of the region with the insulating solution.
$U_{c2}$ is the right border of the region with the metallic solution.
Coexistence region has both solutions. 
 The insulating solution is stable to the right of the curve $U_c$, whereas
the metallic solution is stable on its left.
The empty circle indicates the critical point separating low and high temperature
regions with the first and the second order phase transitions between
metallic and insulating solution respectively.
$U_{c1}$ and $U_{c2}$ curves for different sizes
of the reference system(N=2,4,6) are shown in the inset. }
\end{figure}

 In contrast to the DMFT, where the local Green's function of the reference
system is identified with the one of the original system, we have to
distinguish the two in the DIA formalism. The difference between DMFT and
DIA procedures is best seen from the stationarity condition of
the grand potential functional which results in the identity
%
%
\begin{eqnarray}        %
\sum_{\omega_n}\sum_{n,m}\left[\frac{1}{\vc{G}^{-1}_0-\vc{
\Sigma}(t')}-\vc{G}'\right]_{mn}
\left(\frac{\partial\vc{\Sigma}(\vc{t}')}{\partial\lambda}\right)_{nm}=0.
\label{eq:Euler}
\end{eqnarray}          %
%
%
 The self-energy in DIA is local ($\Sigma_{n,m}=\delta_{n,m}\Sigma_n$). This
implies that the stationarity is automatically fulfilled if the Green's
function of the impurity site in the reference system is identical to 
the local Green's function of the original system. The identity of
the two Green's functions in DIA formalism is expected only when the number
of additional bath sites is large.
 In general, the local Green's function of the original system and the impurity
Green's function of the reference system are related to each other
by the self-energy which they share, but otherwise they are different. Thus
to satisfy the
stationarity condition we also have to take into account the derivative of
the self-energy with respect to a parameter of the reference system
appearing in Eq.~(\ref{eq:Euler}).
In order to facilitate the search for a stationary point, we
reduced the parameter space using particle-hole transformation which is a
symmetry transformation of the Hubbard model at half-filling.
%
%
%
\begin{table}[h]                %
\caption{\label{tab:symetries}\small Constraints on the parameters of the
reference
system imposed by the particle-hole symmetry at half-filling
($\epsilon_1\equiv\epsilon_c, \epsilon_{l\geq 2}\equiv\epsilon_{a,l}$). N is the size
of the reference system. }
\begin{ruledtabular}
\begin{tabular}{ll}
N & Variables and their relations \\ \hline
2 & a) $V_2$   \\
3 & a) $V_2$,$V_3$,~$\epsilon_2=\epsilon_3=\mu$\\
  & b) $V_3=V_2$,$\epsilon_3=2\mu-\epsilon_2$\\
4 & a)
$V_2$,$V_3$,$V_4$,$\epsilon_2=\epsilon_3=\epsilon_4=\mu$\\
  & b)
$V_2$,$V_4=V_3$,$\epsilon_4=2\mu-\epsilon_3$,$\epsilon_2=\mu$\\
5 & a)
$V_2,V_3,V_4,V_5$,$\epsilon_2=\epsilon_3=\epsilon_4=\epsilon_5=\mu$\\
  & b)
$V_2,V_3$,$V_5=V_4$,$\epsilon_2=\epsilon_3=\mu$,$\epsilon_5=2\mu-\epsilon_4$\\
  & c)
$V_3=V_2$,$V_5=V_4$,$\epsilon_3=2\mu-\epsilon_2$,$\epsilon_5=2\mu-\epsilon_4$\\
6 & a)
$V_2,V_3,V_4,V_5,V_6$,$\epsilon_2=\epsilon_3=\epsilon_4=\epsilon_5=\epsilon_6=\mu$\\
  & b)
$V_2,V_3,V_4,V_6=V_5$,$\epsilon_2=\epsilon_3=\epsilon_4=\mu$,$\epsilon_6=2\mu-\epsilon_5$\\
  & c)
$V_2,V_4=V_3,V_6=V_5$,$\epsilon_2=\mu$,$\epsilon_4=2\mu-\epsilon_3$,$\epsilon_6=2\mu-\epsilon_5$\\
\end{tabular}
\end{ruledtabular}
\end{table}                     %
%
 This transformation is defined by the identity 
$P c_{i,\sigma} P^{-1}=e^{i\phi_j}c_{i,\sigma}^\dagger$, 
where $\phi_i$'s are chosen such that $\phi_i-\phi_j=\pi$ and $i,j$ are site
indices. We can choose $\phi=0$ on the sites with even Manhattan distance
and $\phi=\pi$ for the remaining sites. The same transformation is performed
on the impurity site. For the bath sites we have chosen such a particle-hole
transformation that the resulting number of variational parameters is
maximal.
 For the case of the impurity site with $\phi=0$, bath sites
should be transformed according to the expression
$P a_{i,\sigma} P^{-1}=-a_{i,\sigma}^\dagger$. The transformed $H^{SIAM}$ 
Hamiltonian is then
%
%
\beeqa                  %
\label{eq:H_transformed}
PHP^{-1}&&=\sum_{\sigma}(\mu-\epsilon_{c}-U)c_{\sigma}^\dagger
	c_{\sigma}+U n_{\uparrow}n_{\downarrow}\nonumber\\
	&&\hspace{-1.5cm}+2\sum_l(\epsilon_l-\mu)+U+2(\epsilon_c-\mu)
	+\sum_{l,\sigma}(\mu-\epsilon_{a,l}) a_{l,\sigma}^\dagger a_{l,\sigma}\nonumber\\
	&&\hspace{-1.5cm}+\sum_{l,\sigma}V_{l}(c^\dagger_{l,\sigma}a_{l,\sigma}
	+a^\dagger_{l,\sigma}c_{l,\sigma}).
\eneqa                  %
%
  Under the condition that the transformed
Hamiltonian and the original Hamiltonian of the single impurity
Anderson model (SIAM) are the same\cite{Beni}, we
  can put constraints on the parameters of the reference system. Since
$\mu=U/2$, 
we find immediately that $\epsilon_c=0$. Bath states transform 
according to the equations:
%
%
\beeqa                  %
\mu-\epsilon_{a,l}=\epsilon_{a,l'}-\mu,& \quad V_l=V_{l'}.
\label{eq:ph_restrictions}
\eneqa                  %
%
 Combinations following from Eq.~(\ref{eq:ph_restrictions}) are listed in
 Table~\ref{tab:symetries} for different sizes of the reference system.
 It can be shown that the effective size of the reference system is determined
by the number of different $\epsilon$'s, on-site energies of the bath
orbitals. Namely, for the orbitals with $\epsilon$'s
equal to the chemical potential it is possible to perform a unitary
transformation that transforms uncorrelated sites with $\epsilon_i=\mu$ into one
site coupled to the impurity site and the rest of the uncorrelated sites
uncoupled from the impurity site.
 For example, if we had two uncorrelated
sites in the reference system with $\epsilon_{2/3}=\mu$ then it can be shown
that the Hamiltonian can be mapped to the one with the hopping amplitudes
$\tilde{V}_2=\sqrt{V_2^2+V_3^2}$, $\tilde{V}_3=0$. Thus the effective reference system 
has been reduced to a system with only one site. In the case of more bath
sites with $\epsilon_i=\mu$ we can iterate the transformation for two sites
and show that the resulting effective amplitude is
$\tilde{V}_2=\sqrt{\sum_{i=2}^{N}V_i}$, $\tilde{V}_{i>2}=0$. The solutions
in this work correspond to the cases 2a, 4b and 6c from
Table~\ref{tab:symetries} evaluated for the semicircular free-particle
density of states.
%
%
\begin{figure}[htb]
\includegraphics[height=6.0cm]{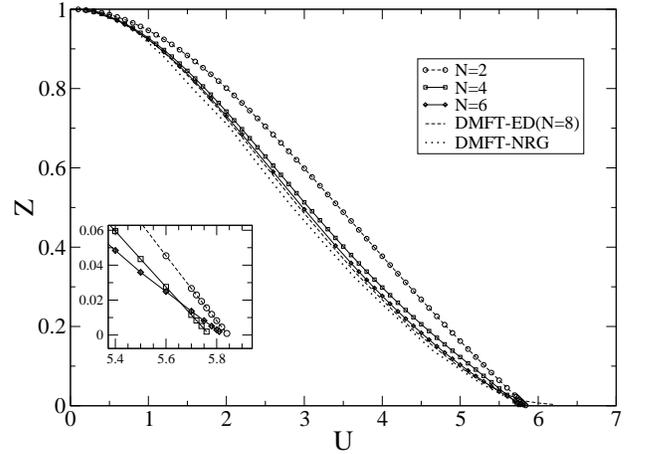}
\vspace*{0.3cm}
\caption{\label{fig:Quasiparticle_weight} 
Quasi-particle weight at T=0 calculated in DIA for different sizes of the
reference system (N=2,4,6). Curves for (N=2,4) have already been
presented\cite{Potthoff0103}. For comparison, DMFT results obtained using
numerical renormalization group(NRG)\cite{Bulla1999} and exact
diagonalization(ED)\cite{Potthoff0103} as impurity solvers are included.}
\end{figure}
%
%
 In Fig.~(\ref{fig:Quasiparticle_weight}) we show the calculated
quasi-particle weight
$Z=1/(1-\partial\Real\Sigma(\omega+i0^+)/\partial\omega)|_{\omega=0}$. Regarding the
interval of U values as a whole we can notice that the convergence of Z with
increasing system size is uniform. That is not true in the vicinity
of the phase transition. Even though the convergence is
not uniform in that region, it seems to be fast. The comparison with DMFT
calculation using exact diagonalization as impurity solver shows a close
connection of the two methods.
 The phase diagram of the paramagnetic state of the Hubbard model in the U-T
plane is shown in Fig.~(\ref{fig:DIA_convergence}).
 Three different regions in temperatures below the critical point, denoted by
an empty circle, can be
distinguished: (a) metallic phase for small values of U, (b) insulating phase
for large U, (c) coexistence of both phases in a triangle-like shape for the
intermediate Coulomb interaction. Whereas the transition below the
critical point is of the first order, we find the crossover from the metallic to the insulating solution in temperatures above the critical
point.
Convergence of the phase diagram for different sizes of the reference system
is shown in the inset of Fig.~(\ref{fig:DIA_convergence}).
%
%
%
%
 On the metallic side of the metal-insulator transition in the DMFT
formalism, the central role is played by a three peak structure in the
spectral function, the middle
peak corresponding to the Abrikosov-Suhl resonance in the impurity model and two
Hubbard bands. That distribution of the spectral weight together with
Table~\ref{tab:symetries} can explain the convergence trends in the solutions
found for different sizes of the reference system. For $N=2$ we have only
one pole in the inverse of the free Green's function of the reference system
at the Fermi level. From the DMFT equation for the semicircular density of
states $
G^{-1}_{0,\sigma}(\omega_n)=i\omega_n+\mu-t^2G_{\sigma}(\omega_n)$ it would
follow that the local Green's function has only one pole on the Fermi level
if the equation holds for each number of bath sites. In the DIA
formalism the connection between the on-site Green's function and the
inverse of the free Green's function in the reference system of a finite
size is more involved, but we believe that rapid convergence of the DIA
results to those obtained in DMFT allows us to use the DMFT equation for the
argumentation. This means that, with N=2, we cannot properly account for the
Hubbard bands central to the insulating phase. The $U_{c1}$ curve is thus
substantially underestimating the extension of the insulating phase region in
comparison with $N=\{4,6\}$ results. The same reasoning explains why
$N=\{3,5\}$ reference systems show no improvement of the metallic solution
with respect to the reference systems with a smaller but even number of sites.
A reference system with an odd number of sites, due to the particle-hole
symmetry, has no pole in the inverse of the free Green's function at
the Fermi level or it has two poles at the Fermi level which can be mapped to
one.
%
%
%
\begin{figure}[ht]
\includegraphics[height=6.0cm]{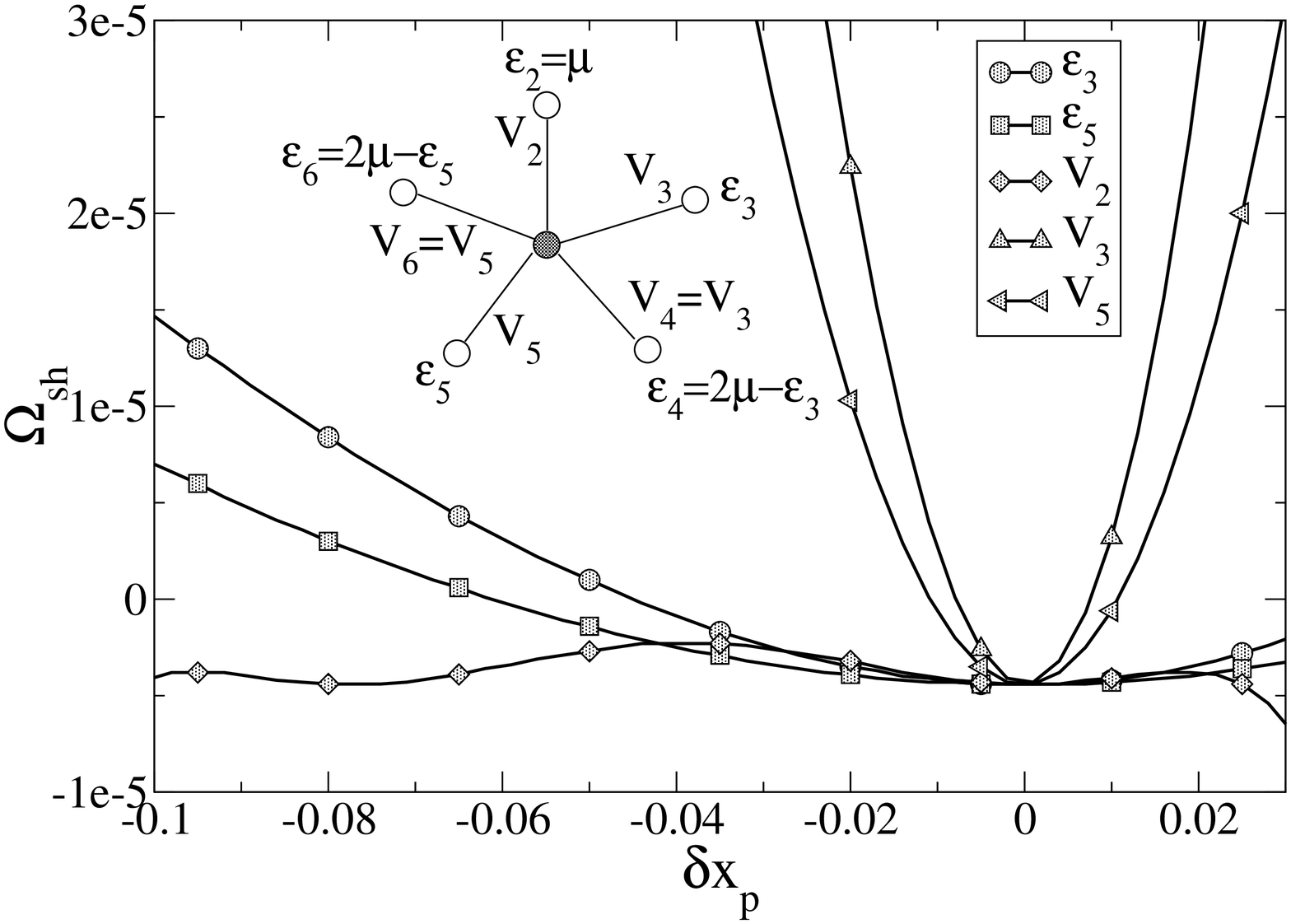}
\vspace*{0.3cm}
\caption{\label{fig:Stationarity}
 Shifted grand potential functional in the direction of the parameters of
the reference system around the insulating solution for T=0.016 and U=4.77.
Grand potential functional is given by 
$\Omega=\Omega_{sh}+2.50717$. Schematic configuration of the
reference system with the corresponding parameters is shown in the upper
left corner.} 
\end{figure}
%
For $N=4$ we can account for the side bands, and $N=6$ brings only a slight
change to the phase diagram.
 Boundaries in the phase diagram are defined by the disappearance of a
stationary point in the parameter space for either metallic or insulating
solution.
 An example of the parameter space is shown in
Fig.~(\ref{fig:Stationarity}). The calculation was done for six atoms. At
half-filling and in the paramagnetic state there are
five independent parameters (upper left corner of the figure). The parameter space is shown in
the region around the stationary point for the insulating phase and U=4.77,
T=0.016.
 Insulating solution disappears when the stationarity condition is not any
more fulfilled in the direction $V_2$.
 As already argued, $V_2$ is related to the weight of the Green's
function of the original system at the Fermi level.
%
%
\begin{figure}[htb]
\includegraphics[height=6.0cm]{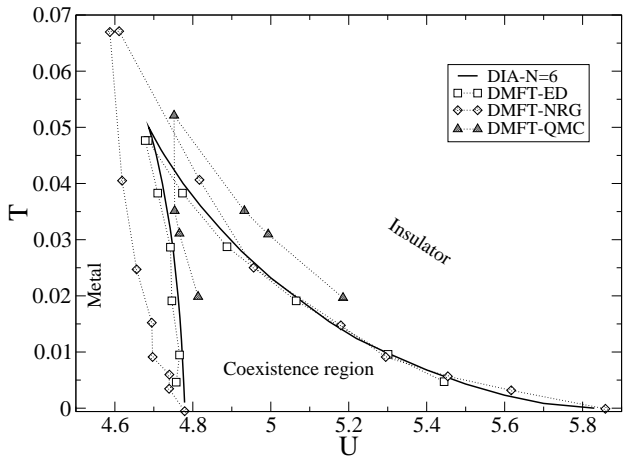}
\caption{\label{fig:MI_transition} 
Metal-insulator transition in the half-filled Hubbard model. The solid line shows
the result from DIA with a reference system of the size N=6. DMFT results with
ED(squares)\cite{Tong2001},NRG(diamonds)\cite{Bulla} and quantum Monte Carlo(triangles)\cite{Joo} as impurity solver are shown for comparison. }
\end{figure}
 Another comparison of the calculations done in the DMFT framework with the
DIA calculation is shown in Fig.~(\ref{fig:MI_transition}) for the whole phase diagram.
 As already noticed for T=0, the DIA calculation shows strong resemblance to
the DMFT-ED results. 
%
%
%
%
%
%
%
%
\par
 In conclusion, we showed how additional information from the reference
system can be used to increase the precision of the calculation in the context
of DIA. Comparisons with the results obtained in DMFT demonstrate close
connection between the DIA and DMFT-ED procedures. The advantage of the DIA
formulation over the DMFT-ED formulation is particularly clear for the case when
the number of bath sites is not sufficiently large to reproduce the local
Green's function of the original system\cite{Potthoff0103}.
 Already for $N=6$, DIA and DMFT-ED give almost the same result for the
metal-insulator transition in 
the paramagnetic state of the Hubbard model at half-filling.
\par
 I would like to thank Prof. C. Gros for his help in a part of this work,
 M. Potthoff for kindly providing additional information on the SFA
prior to the publication, and W. Krauth for a stimulating discussion.
%
%
%
%
%
%
%
\begin{appendix}
%
%
%
%
\section{\label{sec:Green} Derivative of the Green's function}
 In Eq.~(\ref{eq:tr_step_func}) and all subsequent ones we evaluate the
real part of the self-energy and the Green's function on the real axis
(poles are excluded). Thus we will be interested only in the derivatives of the
real parts of those functions.
 Since Green's functions are calculated from their Lehmann representation,\hfill
%
%
\beeqa					%
\label{eq:green_R}
\Real G_{ab}(\omega)&=&\frac{1}{N_d}\sum_{i=1}^{N_d}
\sum_n\left[\frac{\langle n|c_a^\dagger|n_g^{(i)}\rangle
				  \langle n|c_b^\dagger|n_g^{(i)}\rangle}
			{\omega-(E_n-E_0)}\right.\nonumber\\
			&+&\left.\frac{\langle n|c_b|n_g^{(i)}\rangle
					\langle n|c_a|n_g^{(i)}\rangle}
			{\omega-(E_0-E_n)}
		\right],
\eneqa					%
%
%
%
%
%
 where $N_d$ is the degeneracy of the ground state, $|n_g^{(i)}\rangle$
are degenerate ground-state eigenfunctions and $E_0$ is the ground-state energy,
 the derivative with respect to $\omega$ is straightforward to evaluate. 
 Summation over $n$ extends over all eigenstates. In Eq.~(\ref{eq:green_R})
we used the fact that the matrix representation of the Hamiltonian and the
corresponding eigenfunctions will be real, and organized all matrix elements
so that the ground states are in the ket state.
 \par
 The derivative of the Green's function with respect to a parameter of the
reference system $\lambda\in\{\epsilon_i,V_i\}$ can be calculated by applying the derivative to its Lehmann
representation.
 For each pair of indices $(i,n)$ there are two terms contributing to the
Green's function, where the first term is the projection of the state
created out of the N-particle ground state upon addition of a particle, on
the eigenstates of the (N+1)-particle system. The second term 
is the projection of the ground state with added hole. 
 We can get the second term 
by performing the following replacements in the first term: $c_a^{\dagger}\Rightarrow c_a$,
$c_b^{\dagger}\Rightarrow c_b$, and by exchanging energies ($E_0\Leftrightarrow
E_n$) in the denominator.
 The derivative of the first term is
%
%
\beeqa					%
\frac{\partial}{\partial\lambda} &&
\frac{\langle n|c_a^\dagger|n_g^{(i)}\rangle
				  \langle n|c_b^\dagger|n_g^{(i)}\rangle}
			{\omega-(E_n-E_0)} \nonumber\\
&&=\frac{\frac{\partial\langle n|c_a^\dagger|n_g^{(i)}\rangle}{\partial\lambda}
		\langle n|c_b^\dagger|n_g^{(i)}\rangle
		+\frac{\partial\langle n|c_b^\dagger|n_g^{(i)}\rangle}{\partial\lambda}
		\langle n|c_a^\dagger|n_g^{(i)}\rangle
  		}{\omega-(E_n-E_0)}
\nonumber\\
&&+\frac{\langle n|c_a^\dagger|n_g^{(i)}\rangle\langle n|c_b^\dagger|n_g^{(i)}\rangle
		}{[\omega-(E_n-E_0)]^2}
		\left(\frac{\partial E_n}{\partial\lambda}-\frac{\partial E_0}{\partial\lambda}
		\right).
\label{eq:g_deriv_inte}
\eneqa					%
%
%
 In the term with $\omega-(E_n-E_0)$ in the denominator, the second part
in the numerator is obtained from the first by changing
$c_a^\dagger$ to $c_b^\dagger$. The remaining derivatives are:
%
%
\beeqa					%
\frac{\partial\langle n|c_a^\dagger|n_g^{(i)}\rangle}{\partial\lambda}&=&
\frac{\partial\langle n|}{\partial\lambda}c_a^\dagger|n_g^{(i)}\rangle
+\langle
n|c_a^\dagger\frac{\partial|n_g^{(i)}\rangle}{\partial\lambda},\nonumber\\
\frac{\partial E_n}{\partial\lambda}
&,&
\frac{\partial E_0}{\partial\lambda}
\eneqa					%
%
%
%
%
 In the end we have the problem of determining derivatives of 
eigenvectors and eigenvalues with respect to the parameter $\lambda$.
 It is well known that the terms up to the second order 
in the Rayleigh-Schr\"odinger perturbation theory correspond to the order of perturbation
$\lambda$. We can thus use the perturbation to
determine derivatives of eigenvalues and eigenvectors.
Special attention should be paid to the
case when some eigenstates are degenerate. If we denote the perturbation
as $H'=\lambda V$, then,  in the perturbation theory\cite{Baym}, we have to use states from
the subspace which diagonalize the operator $V$. Assuming that the
above requirement is fulfilled, it follows that
%
%
\beeqa					%
\frac{\partial\langle n|c_a^\dagger|n_g^{(i)}\rangle}{\partial\lambda}
&=&\sum_{\substack{m\\E_m\neq E_0}}
\frac{\langle m|V|n_g^{(i)}\rangle\langle n|c_a^\dagger|m\rangle}
{E_0-E_m}\nonumber\\
&+&\sum_{\substack{n'\\E_{n'}\neq E_n}}
\frac{\langle n'|c_a^\dagger|n_g^{(i)}\rangle\langle n'|V|n\rangle}
{E_n-E_{n'}}
\eneqa					%
%
%
 and
%
%
\beeqa					%
\frac{\partial E_n}{\partial\lambda}
=\langle n|V|n\rangle
,\qquad
\frac{\partial E_0}{\partial\lambda}                                  
=\langle n_g^{(i)}|V|n_g^{(i)}\rangle
\eneqa					%
%
%
 After some rearrangements, the derivative of the Green's function has
the form
%
%
\beeqa					%
&&
\hspace{-1.2cm}\frac{\partial\Real G_{ab}(\omega,\lambda)}{\partial\lambda}
= \frac{1}{N_d}\sum_n\sum_{i=1}^{N_d}
				\nonumber\\
				&&\hspace{0.0cm}
					 \left\{\frac{A_{i,n}^{ab}}
						{\omega-(E_n-E_0)}
						+\frac{B_{i,n}^{ab}}{\omega-(E_0-E_n)}
					 \right.
					 \nonumber\\
				&&\hspace{0.0cm}
						\left.
						+\frac{C_{i,n}^{ab}}
								{[\omega-(E_n-E_0)]^2}
						+\frac{D_{i,n}^{ab}}
								{[\omega-(E_0-E_n)]^2}
					   \right\},
\eneqa					%
%
%
with weights of the first order poles given by
%
%
\beeqa					%
\hspace*{-2cm}
A_{i,n}^{ab}&=&\sum_{\substack{n'\\E_{n'}\neq E_n}}
				\frac{\langle n'|V|n\rangle}{E_n-E_{n'}}
				\left(\langle n'|c_a^\dagger|n_g^{(i)}\rangle\langle n|c_b^\dagger|n_g^{(i)}\rangle
				\right.\nonumber\\
	 		&+& \left.\langle n|c_a^\dagger|n_g^{(i)}\rangle\langle n'|c_b^\dagger|n_g^{(i)}\rangle
				\right) \nonumber\\
	&&+\sum_{\substack{m\\E_m\neq E_0}}
				\frac{\langle m|V|n_g^{(i)}\rangle}{E_0-E_m}
				\left(\langle n|c_a^\dagger|m\rangle\langle n|c_b^\dagger|n_g^{(i)}\rangle
				\right.\nonumber\\
			&+& \left.\langle n|c_a^\dagger|n_g^{(i)}\rangle\langle n|c_b^\dagger|m\rangle
				\right)  
\eneqa					%
%
%
 and $B_{i,n}^{ab}=A_{i,n}^{ab}(c_a^\dagger\Leftrightarrow
c_a,c_b^\dagger\Leftrightarrow c_b)$, where $\Leftrightarrow$ means the exchange
of the operators. The second order poles have the weights $C_{i,n}^{ab}$
equal to 
%
\beeqa					%
\langle n|c_a^\dagger|n_g^{(i)}\rangle
			\langle n|c_b^\dagger|n_g^{(i)}\rangle
			\left(\langle n|V|n\rangle
				  -\langle n_g^{(i)}|V| n_g^{(i)}\rangle
			\right)
\eneqa					%
%
%
%
 and $D_{i,n}^{ab}=-C_{i,n}^{ab}(c_a^\dagger\Leftrightarrow
c_a,c_b^\dagger\Leftrightarrow c_b)$.

 Along the same lines we can calculate the derivative of the Green's function 
for finite temperatures. For the Green's function $\Real G_{ab}(\omega)$ being given by the Lehmann
representation
%
\beeqa                  %
\label{eq:greenT}
\hspace*{-5mm}
\sum_{n,n'}
\frac{1}{Z}\left(e^{-\beta E_{n'}}+e^{-\beta E_n}\right)
\frac{\langle n'|c_a|n\rangle\langle n|c_b^\dagger|n'\rangle}
{\omega-(E_n-E_{n'})},
\eneqa                  %
%
the derivative $\frac{\partial\Real G_{ab}(\omega)}{\partial\lambda}$
is given by the function
%
\beeqa                  %
\label{eq:deriv_greenT}  
\hspace*{-1cm}
\sum_{n,n'}
\left[\frac{A_{n,n'}^{ab}}{\omega-(E_n-E_n')}
+\frac{B_{n,n'}^{ab}}{\left[\omega-(E_n-E_n')\right]^2}
\right],
\eneqa                  %
%
%
with $A_{n,n'}^{ab}$ equal to
\begin{widetext}
\beeqa					%
&&
\frac{e^{-\beta E_n}+e^{-\beta	 E_{n'}}}{Z}
			\left[\sum_{\substack{l\\E_l\neq E_n}}
				\frac{\langle l|V|n\rangle}{E_n-E_l}
				\left(\langle l|c_a^\dagger|n'\rangle\langle n|c_b^\dagger|n'\rangle
	 		+ \langle n|c_a^\dagger|n'\rangle\langle l|c_b^\dagger|n'\rangle
				\right)
	+\sum_{\substack{m\\E_m\neq E_{n'}}}
				\frac{\langle m|V|n'\rangle}{E_{n'}-E_m}
				\left(\langle n|c_a^\dagger|m\rangle\langle n|c_b^\dagger|n'\rangle
	\right.\right.\nonumber\\
	&&+\left.\left.\langle n|c_a^\dagger|n'\rangle\langle n|c_b^\dagger|m\rangle
				\right)\right]
	-\frac{1}{Z}\left[
	  e^{-\beta E_n}\left(\beta\frac{\partial E_n}{\partial\lambda}
      +\frac{1}{Z}\frac{\partial Z}{\partial\lambda}\right)
      +e^{-\beta E_{n'}}\left(\beta\frac{\partial E_{n'}}{\partial\lambda}
      +\frac{1}{Z}\frac{\partial Z}{\partial\lambda}\right)\right]
	\langle n|c_a^\dagger|n'\rangle\langle n|c_b^\dagger|n'\rangle
\eneqa					%
\end{widetext}
 and
%
%
\beeqa					%
B_{n,n'}^{ab}&=&\frac{1}{Z}
			\left(e^{-\beta E_n}+e^{-\beta E_{n'}}\right)
			\langle n|c_a^\dagger|n'\rangle
			\langle n|c_b^\dagger|n'\rangle
			\nonumber\\
			&&
			\times\left(\langle n|V|n\rangle
				  -\langle n'|V| n'\rangle
			\right).
\eneqa					%
%
%
%
%
%
\section{\label{sec:Self_energy} Self-energy and its derivatives}
 Self-energy $\vc{\Sigma}(\vc{t}',\omega)$ is defined by the Dyson equation
%
%
\beeq					%
\label{eq:self_energy}
{\bf\Sigma}(\omega)={\bf G}_0^{-1}(\omega)-{\bf G}^{-1}(\omega),
\eneq					%
%
 In order to be able to perform high precision calculation, it is necessary
to be able to do exact subtraction of $G_0^{-1}(\omega)$ and $G^{-1}(\omega)$. 
 Direct subtraction of the calculated inverses of $G_0(\omega)$ and
$G(\omega)$ at some point $\omega$ is not a method of choice if we strive
for high precision calculation. The reason lies in the finite precision
arithmetics in each computer. Details of this discussion will be given at
the end of this section.
\par
 An alternative to the subtraction of the evaluated inverses of Green's
functions is to subtract them as functions of the frequency and only
then evaluate them in the frequency we are interested in. The difference
with respect to the subtraction of already evaluated functions is that in
this way we always subtract numbers of order one, whereas 
in the direct subtraction the subtracted numbers are sometimes large (in the
vicinity of a pole) but close to each other.
 In this section we will focus only on the DIA case, i.e., local self-energy.
Generalization to the non-local self-energies is straightforward.
\par
 If we mark the impurity site with index $1$, then the self-energy on that
site has a form
%
%
\beeq					%
\Real {\Sigma_{11}}(\omega)=[{G}_{0,11}(\omega)]^{-1}-[{G}_{11}(\omega)]^{-1}
\eneq					%
%
%
 where, in the case that each bath atom is connected only with the impurity
site (starlike bath), the inverse of the free Green's function is obtained
from the equations of motion and reads
%
%
\beeq					%
[\Real G_{0,11}(\omega)]^{-1}=\omega+\mu-\epsilon_1
		-\sum_{l=2}^{N_b}\frac{V_l^2}{\omega+\mu-\epsilon_l},
\label{eq:g011_inv}
\eneq					%
%
%
 where $\epsilon_l$ is on-site energy, $V_l$ is the hopping amplitude
between bath sites and the impurity site, $N_b$ is the number of atoms in the
reference system and $\mu$ is the chemical potential. 
 From the Lehmann representation we know that the real part of the thermal
one-particle Green's function on the real axis can be written in the form
%
%
\beeq					%
\Real G_{11}(\omega)=\sum_{i=1}^{N_{G}}\frac{C^{(1),G}_i}{\omega-\epsilon_i}
\label{eq:g11}
\eneq					%
%
%
 where ${C^{(1),G}_i}>0$ and $\sum_i{C^{(1),G}_i}=1$, ~$N_{G}$ is a number of
poles and $\epsilon_i$'s are poles of $G_{11}(\omega)$. The inverse of
$\Real {G}_{11}(\omega)$ is then
%
%
\beeq					%
[\Real G_{11}(\omega)]^{-1}=C^{(0),G^{-1}}+\omega
+\sum_{i=1}^{N_{G}-1}\frac{C^{(1),G^{-1}}_i}{\omega-\omega_i},
\label{eq:g11_inv}
\eneq					%
%
%
where $\omega_i$'s are zeros of $G$ (poles of $G^{-1}$). 
Coefficients $C^{(1),G^{-1}}_i$ are obtained if we equate the expression
in Eq.~(\ref{eq:g11_inv}) with $1/\Real G_{11}(\omega)$, where 
$\Real G_{11}(\omega)$ is given by Eq.~(\ref{eq:g11}), and then calculate
the limit $\omega\to\omega_j$. We get
%
%
\beeq					%
C^{(1),G^{-1}}_j=\lim_{\omega\to\omega_j}\frac{\omega-\omega_j}{\sum_{i=1}^{N_{G}}
\frac{C^{(1),G}_i}{\omega-\epsilon_i}}
=\frac{-1}{\sum_{i=1}^{N_{G}}
\frac{C^{(1),G}_i}{(\omega_j-\epsilon_i)^2}}
\label{eq:coeff_GI}
\eneq					%
%
%
 To have a value associated with $C^{(0),G^{-1}}_i$, we once again equate the
expression in Eq.~(\ref{eq:g11_inv}) with $1/\Real G_{11}(\omega)$ from
Eq.~(\ref{eq:g11}) in any non-singular point $\omega$, but this time we use
just calculated $C^{(1),G^{-1}}_j$ coefficients.
 The procedure for the calculation of the derivative of $\Real G^{-1}$ with
respect to some parameter of the reference system $\lambda$ resembles the
one used for the calculation of $\Real G^{-1}$. From Eq.~(\ref{eq:g11_inv}) we see
that $\partial[\Real G_{11}(\omega)]^{-1}/\partial\lambda$ can be written in
the form
%
%
\beeqa					%
\hspace*{-0.6cm}
\frac{\partial[\Real G_{11}(\omega)]^{-1}}{\partial\lambda}&=&
\frac{-1}{[\Real G_{11}(\omega)]^2}\frac{\partial\Real G_{11}(\omega)}{\partial\lambda}
\label{eq:g11_inv_deriv_def}\\
&&\hspace*{-2.8cm}
=C^{(0),dG^{-1}}
+\sum_{i=1}^{N_{G}-1}\frac{C^{(1),dG^{-1}}_i}{\omega-\omega_i}
+\sum_{i=1}^{N_{G}-1}\frac{C^{(2),dG^{-1}}_i}{(\omega-\omega_i)^2}
\label{eq:g11_inv_deriv}
\eneqa					%
%
%
The coefficients $C^{(1),dG^{-1}}_i$ and $C^{(2),dG^{-1}}_i$ follow
 from 
%
%
\beeqa					%
\hspace*{-0.7cm}
C_j^{(1),dG^{-1}}&=&\lim_{\omega\to\omega_j}
	\frac{\partial}{\partial\omega}\left[\frac{-(\omega-\omega_j)^2}{[\Real G_{11}(\omega)]^2}
	\frac{\partial\Real G_{11}(\omega)}{\partial\lambda}\right]
	\nonumber\\
&&\hspace*{-1.5cm}
	=-2 C_j^{(1),G^{-1}}\left(C^{(0),G^{-1}}+\omega_j+\sum_{\substack{i=1\\i\neq j}}
	\frac{C_i^{(1),G^{-1}}}{\omega_j-\omega_i}
	\right)
	\nonumber\\
&&\hspace*{-1.8cm}
	\times\left.\frac{\partial\Real G_{11}(\omega)}{\partial\lambda}
	\right|_{\omega=\omega_j}
	-\left(C_j^{(1),G^{-1}}\right)^2\left.\frac{\partial^2 \Real G_{11}}{\partial\omega\partial\lambda}
	\right|_{\omega=\omega_j}
	\label{eq:g11_inv_deriv_coeff1}
	\\
C_j^{(2),dG^{-1}}&=&\lim_{\omega\to\omega_j}
	\left[(\omega-\omega_j)^2\frac{-1}{[\Real G_{11}(\omega)]^2}
	\frac{\partial\Real G_{11}(\omega)}{\partial\lambda}\right]
	\nonumber\\
	&=&-\left(C_j^{(1),G^{-1}}\right)^2
	\left(\frac{\partial\Real G_{11}(\omega)}{\partial\lambda}
	\right)_{\omega=\omega_j}
\label{eq:g11_inv_deriv_coeff2}
\eneqa					%
%
%
 Unfortunately, the above equation for $C^{(1),dG^{-1}}$ turned out to be
numerically unstable, i.e., in practice we get only a few relevant digits
correctly. There is another form in which we can write it and which
is numerically stable. 
 Taking the derivative of Eq.~(\ref{eq:g11_inv}) with respect to $\lambda$, we see that 
$C^{(1),dG^{-1}}_j=\partial C^{(1),G^{-1}}_j/\partial\lambda$. Now using
Eq.~(\ref{eq:coeff_GI}) we can show that $C^{(1),dG^{-1}}_j$ equals to
%
%
\beeqa					%
	-\left[C_j^{(1),G^{-1}}\right]^2\left(
	\left.\frac{\partial^2\Real G}{\partial\omega\partial\lambda}\right|_{\omega_j}
	+\frac{\partial\omega_j}{\partial\lambda}\left.\frac{\partial^2\Real G}{\partial\omega^2}\right|_{\omega_j}
	\right).
	\label{eq:g11_inv_deriv_coeff1_stable}
\eneqa					%
%
%
 From the definition of $\omega_j$, i.e., zero of the Green's function, its
derivative can be deduced and equals
%
%
\beeqa					%
\frac{\partial\omega_j}{\partial\lambda}=\left.-C_j^{(1),G^{-1}}\frac{\partial
\Real G}{\partial\lambda}\right|_{\omega_j}.
\eneqa					%
%
%
 Coefficient $C^{(0),dG^{-1}}$ then follows from
Eq.~(\ref{eq:g11_inv_deriv}) with the coefficients
Eq.~(\ref{eq:g11_inv_deriv_coeff1_stable}, \ref{eq:g11_inv_deriv_coeff2})
and Eq.~(\ref{eq:g11_inv_deriv_def}) if we equate them in an arbitrary
non-singular point $\omega$.
 Having the inverses of the Green's functions and their derivative expressed as
functions of frequency allows the representation of the self-energy 
with a sum of poles added to a constant term. 
\begin{table}[htb]
\caption{\label{tab:sigma_precision_degradation} Illustration of the
degradation in precision by the evaluation of the self-energy for different
frequencies using evaluated
inverses of the Green's functions in Eq.~\ref{eq:self_energy} ($2^{nd}$ column) 
in comparison with the result obtained from 
Eq.~\ref{eq:sig_analytic} ($3^{rd}$ column).
 The calculation is done for the single impurity Anderson model with one bath 
site and for the parameters: U=4, $\mu=U/2$, $\epsilon_c=0$, $T=0.001$,
$\epsilon_2=\mu$,
$V_2=0.5$. }
\begin{ruledtabular}
\begin{tabular}{c|c|c}
$\omega$ & numeric & analytic \\
\hline
                   -1e-07&         2.00278960194782&         2.00000017777778\\
                   -5e-08&         2.01115762927247&         2.00000008888889\\
                   -1e-08&         2.27893444799702&         2.00000001777778\\ 
                   -4e-09&         3.74339942335428&         2.00000000711111\\ 
                   -2e-09&         8.97438195502764&         2.00000000355556\\
                  7.7e-23&         1.6761776955e+15&                        2\\
                    2e-09&         8.97277702135762&         1.99999999644445\\
                    4e-09&         3.74317856255584&         1.99999999288889\\
                    1e-08&         2.27895191107018&         1.99999998222222\\
                    5e-08&         2.01115782492434&         1.99999991111111\\
                    1e-07&         2.00278900765716&         1.99999982222222\\
\end{tabular}
\end{ruledtabular}
\end{table}
Causality of $\Sigma$ demands that each pole has a positive residue.
As a consequence, the poles of $[G_{0,11}(\omega)]^{-1}$ in Eq.~(\ref{eq:g11}) 
are compensated by the poles of $[G_{11}(\omega)]^{-1}$ from 
Eq.~(\ref{eq:g11_inv}). 
The final expression
for the self-energy can thus be written in the form
%
%
\beeq					%
\Real \Sigma_{11}(\omega)=C^{(0),\Sigma}
+\sum_{i=1}^{N_G-1}\frac{C^{(1),\Sigma}_i}{\omega-\omega_i},
\label{eq:sig_analytic}
\eneq					%
%
%
 where $C^{(0),\Sigma}=\mu-\epsilon_1-C^{(0),G^{-1}}$ and 
$C^{(1),\Sigma}_i=-\sum_{l}\delta_{\epsilon_l,\omega_i}
V_l^2-C_i^{(1),G^{-1}}$. 
Analytic form of the derivative of the self-energy
$\frac{\partial\Real \Sigma_{11}(\omega)}{\partial\lambda}$ is determined 
by Eqs.~(\ref{eq:sig_analytic},\ref{eq:g011_inv},\ref{eq:g11_inv_deriv}) 
%
%
\beeq					%
C^{(0),d\Sigma}
+\sum_{i=1}^{N_G-1}\frac{C^{(1),d\Sigma}_i}{\omega-\omega_i}
+\sum_{i=1}^{N_G-1}\frac{C^{(2),d\Sigma}_i}{(\omega-\omega_i)^2}
\label{eq:dsig_analytic}
\eneq					%
%
%
 with 
%
\beeqa					%
C^{(0),d\Sigma}=-\frac{\partial\epsilon_1}{\partial\lambda}-C^{(0),dG^{-1}},
\eneqa
\beeq
C^{(1),d\Sigma}_i=-2\sum_l\delta_{\epsilon_l,\omega_i}V_l\frac{\partial
V_l}{\partial\lambda}
   -C^{(1),dG^{-1}}_i,
\eneq
\beeqa
C^{(2),d\Sigma}_i=-\sum_l\delta_{\epsilon_l,\omega_i}V_l^2
   \frac{\partial\epsilon_l}{\partial\lambda}-C^{(2),dG^{-1}}_i.
\eneqa					%
%
%
%
 The derivative of the self-energy with respect to the frequency $\omega$
is immediately evaluated from Eqs.~(\ref{eq:g011_inv},\ref{eq:g11_inv})
and reads
%
%
\beeq					%
\frac{\partial\Real\Sigma_{11}(\omega)}{\partial\omega}=
-\sum_{i=1}^{N_G-1}
\frac{C^{(1),G^{-1}}_i+\sum\limits_lV_l^2\delta_{\epsilon_l,\omega_i}}
{(\omega-\omega_i)^2}.
\label{eq:dsig_omega_analytic}
\eneq					%
%
%
 As commented at the beginning of this section, the above scheme is
numerically more stable than the direct subtraction of the functions
evaluated at a certain frequency $\omega$. This is illustrated in
Table~\ref{tab:sigma_precision_degradation}.
 With a given precision of 8-byte floating point number (about 15 relevant
digits) we see that direct evaluation gets the first relevant
digit wrong already when the frequency is at a distance of $\approx 10^{-8}$
from a spurious pole. 
Instead of having a finite value at $\omega=0$, numerical
calculation diverges. The non-existing divergence in the subtraction of the
evaluated inverses of Green's functions has been removed in the step
where we subtracted coefficients of $G_0^{-1}$ and $G^{-1}$. As a result we
have a small weight of the self-energy. If we had not nullified all
small weights (small in comparison with the strongest pole), we would get
very weak divergence in that point, but we know from the evaluation of the trace
terms that weak poles give a small contribution to the grand
potential. This allows us to nullify weights smaller than the precision
determined primarily by the integration performed in the $\Tr\ln(-G_k)$ term.
Aside from the loss of precision in the numerical algorithm, the appearance of the
singularity at a point where the self-energy is supposed to be analytic might
cause problems to the zero-searching algorithm.
\end{appendix}
\end{document}